# Stream State-tying for Sign Language Recognition

Jiyong Ma[1*], Wen Gao[1,2], and Chunli Wang[3]

[1]Institute of Computing Technology, Chinese Academy of Science, Beijing 100080, China

[2] Department of Computer Science, Harbin Institute of Technology, Harbin, China

[3] Department of Computer Science, Dalian University of Technology, Dalian, China

**Abstract**

In this paper, a novel approach to sign language recognition based on state tying in each of data streams is presented. In this framework, it is assumed that hand gesture signal is represented in terms of six synchronous data streams, i.e., the left/right hand position, left/right hand orientation and left/right handshape. This approach offers a very accurate representation of the sign space and keeps the number of parameters reasonably small in favor of a fast decoding. Experiments were carried out for 5177 Chinese signs. The real time isolated recognition rate is 94.8%. For continuous sign recognition, the word correct rate is 91.4%.

*Keywords*: Sign language recognition; Automatic sign language translation; Hand gesture recognition; Hidden Markov models; State-tying; Multimodal user interface; Virtual reality; Man-machine systems.

## 1. Introduction

Sign language is one of the natural means of exchanging information for the hearing impaired. It is a kind of visual language via hand and arm movements accompanying facial expression and lip motion. The facial expression and lip motion are less important than hand gestures in sign language, but they may help to understand some hand gestures. Digitized devices can be used to measure the temporal and spatial information of hand gestures, the typical devices are data gloves, position trackers. In this paper, we use two CyberGloves and a position tracker, i.e., Pohelmus 3SPACE with two receivers positioned on the wrist of each CyberGlove and one fixed at thorax as input devices to measure gestures.

Chinese sign language is classified into two categories. One is hand gesture in which each gesture corresponds to a Chinese phrase. The other is fingerspelling in which each alphabet corresponds to a posture, and each Chinese sign corresponds to several postures performed continuously. Hand gestures are often used for communication among the hearing impaired in the daily life. The Chinese sign language books contained about 5500 conventional Chinese phrases, in which each corresponds to a gesture or a sign in sign language. This means that basic unit in Chinese Sign language is the Chinese phrase.

Sign language recognition has many applications, such as controlling the motion of a human avatar in a virtual environment (VE) via hand gesture recognition [1], multimodal user interface in virtual reality (VR) system [2].

Attempts to automatically recognize sign language began to appear in literature at the end of the 80's. For the case of posture recognition, the recognition algorithms include inductive algorithms such as ID3, NewID, C4.5, CN2, and HCV, RIEVL [3], neural network approach such as the hybrid approach of radial basis functions (RBF), inductive decision trees [4] and a fuzzy min-max neural network [5]. Fels and Hinton's [6] and Fell's [7] developed a system using a VPL DataGlove Mark II with a Polhemus tracker as input devices. In this system, the neural network was employed for classifying hand gestures. Takahashi and Kishino [8] investigated understanding the Japanese Kana manual alphabets corresponding to 46 signs using a VPL DataGlove. The system could correctly recognize 30 of the 46 signs, while the remaining 16 could not be reliably identified. Murakami and Taguchi [9] made use of recurrent neural nets for sign Language recognition. They trained the system on 42 handshapes in the Japanese finger alphabet using a VPL Data Glove. The recognition rate is 98 per cent. W.Kadous [10] demonstrated a system based on Power Gloves to recognize a set of 95 isolated Auslan signs with 80% accuracy using fast match methods. Tung and Kak [11] described automatic learning of robot tasks through a DataGlove interface. Kang and Kikuchi [12] designed a system for simple task learning by human demonstration. Kisti Grobel and Marcell Assan [13] used HMMs to recognize isolated signs with 91.3% accuracy out of a 262-sign vocabulary. They extracted the features from video recordings of signers wearing colored gloves. Charaphayan and Marble [14] investigated a way using image processing to understand American Sign Language (ASL). This system can recognize correctly 27 of the 31 ASL symbols.

For the case of continuous sign recognition, Sterner [15] reported that a color camera was used, and the users wore a yellow glove on their right hand and orange one on their left. In this case, a correct rate was achieved by 91.3 per cent. By imposing a strict grammar on this, it was shown that accuracy rates for 40 signs more than 99 per cent were possible with real-time performance.

R.H.Liang and M.Ouhyoung [16] used HMM for continuous recognition of Taiwan Sign language with a vocabulary between 71 and 250 signs based data gloves as input devices. However, the system required gestures performed by the signer slowly enough to detect the word boundary. This requirement is hardly ensured for practical applications. C.Vogler and D.Metaxas [17] used HMMs for continuous ASL recognition with a vocabulary of 53 signs and a completely unconstrained sentence structure. C.Vogler and D.Metaxas [18] described an approach to continuous, whole-sentence ASL recognition that uses phonemes instead of whole signs as the basic units. They experimented with 22 words and achieved similar recognition rates with phoneme and word-based approaches.

By reviewing of foregoing research work we know that most research on sign language recognition were made on small test vocabulary. For large vocabulary recognition, Ho-Sub Yoon recently [19] reported that a sign vocabulary consisting of 1,300 alphabetical gestures were recognized using HMMs. C.Vogler and D.Metaxas [20] pointed that the major challenge to sign language recognition is how to develop approaches that scale well with increasing vocabulary size. He used parallel HMMs (PaHMMs) [21] to solve the problem. PaHMMs process multi streams independently. Thus, they can also be trained independently, and do not require consideration of the different combinations at training time. He ran several experiments with a 22-sign vocabulary and demonstrated that PaHMMs could improve the robustness of HMM-based recognition even at a small scale. It would be possible to permit some stream asynchrony. When streams are not frame synchronous, the complexity that the decoding algorithm required may be considerably greater than that for a standard recognizer. Results to date in speech recognition have indicated that allowing asynchrony among streams does not give any signification performance improvement [22].

State tying is a popular and efficient technique developed in speech recognition community. In this paper, we will address the scalability problem in large vocabulary sign recognition with stream state tying. The size of sign vocabulary is 5177 signs, and we believe this is the largest test scale in sign language recognition community up to date.

The organization of this paper is as the following. Section 2 addresses basic units in hand gestures. Section 3 presents the state tying in each of streams. Section 4 describes a fast algorithm for isolated sign recognition. Section 5 describes the continuous sign recognition including movement epenthesis models and a fast-matching approach. Section 6 summarizes the search algorithm in sign language recognition. Section7 contains the experiment results. Section 8 contains the conclusion.

## 2. Basic units in hand gestures

Compared with human languages such as speech and handwriting, sign language also should have its own basic units. The aim to search the basic units in sign language is to build scalable systems with respect to large vocabulary size. The basic units in speech are phonemes, and the basic units in handwriting are the alphabets or strokes, etc. In contrast, the basic units in sign language are more complex. From the temporal and spatial analysis, for each time instant, hand shape, hand position and hand orientation are three measurable factors forming a hand spatial unit in whole sign space, we call it *a spatial unit* of a hand gesture. In addition, a hand gesture has a start position that corresponds to a spatial unit, transitional segment, and a stop position that also corresponds to a spatial unit. Therefore, a hand gesture can be viewed as a trajectory in a high dimensional space.

From the analysis above the basic spatial units in hand gestures should include the following six data streams:

right handshape, right hand position, right hand orientation, left handshape, left hand position and left hand orientation. And they forms a vector denoted by （$X_1(t),X_2(t),X_3(t),X_4(t),X_5(t),X_6(t)$）.Each data stream is also a vector. The six data streams are observable and synchronous. The feature extraction approach for independent singer positions can be found in our previous work [24,27,28]. In Fig.1 ($X_1,X_2,X_3$) and ($X'_1,X'_2,X'_3$) denote the coordinates of an object point with respect to the Cartesian coordinate systems of the transmitter and receiver ,respectively. $S$ is the position vector of the receiver with respect to Cartesian coordinate systems of the transmitter.

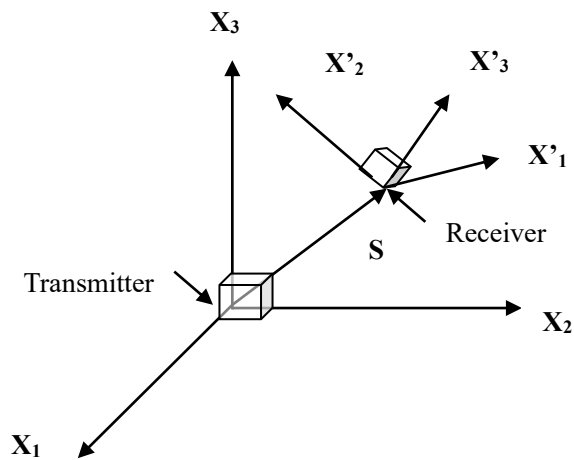

**Figure 1.** The relationships between the transmitter and receivers

Two CyberGloves with 18 sensors and a Pohelmus 3-D tracker are used as input devices in our system. The 18 sensors in each CyberGlove include two bend sensors on each finger, four abduction sensors, plus sensors measuring thumb crossover, palm arch, wrist flexion and wrist abduction.

As there are noises in measurement, and the data are different even for the same gesture performed in different time. Therefore, the vector is a random vector.

From the distinguishable ability of human vision point of view the number of different patterns in each data stream should not be very large so that human can recognize them easily. For streams such as handshapes, hand positions and hand orientations, the number of different patterns in each stream is usually not larger than 1024. But the number of their combinations in the whole sign space is very large. As C.Vogler and D.Metaxas [20] pointed that the number of possible combinations of different patterns in streams after enforcing linguistic constraints is approximately $5.510^8$. It seems impossible to obtain a few spatial basic units in the whole sign space. Therefore, the basic units should be found in each of data streams.

## 3. Tying

The aim of tying is to reduce computation load in decoding process. The tying can be at different levels.

Sharing the same HMM among some different HMMs is named as *HMM tying in the whole sign space*. This is the first level tying. Clustering different HMMs is more difficult than clustering different state models. The reason is that different HMMs may have different topologies, different transition probabilities and different state observation densities. So, the measure to distinguish different HMMs is difficult to choose.

Sharing the same state models among some different state models is named as *state tying in the whole sign space.* This is the second level tying. For this problem a measure to cluster the state models can be defined so that some state models are tied. Computation load of state models can be reduced using this approach, and the state tying approach [24] developed in speech recognition community can be used.

There are a lot of possible ways to cluster HMM models or state models in speech recognition. These approaches can be directly used for sign language recognition. However, tying at HMM level and state level for sign language recognition is not as effective as that for speech recognition. The reason has been descried in detail in section 2. Therefore, tying in each data stream or in sign subspace is proposed, where the data streams include hand shape, hand position and hand orientation. This is the third level tying. It is named as *stream state tying.* This level tying is more efficient for calculating state observation probabilities for sign language recognition. Unlike state tying in speech recognition the stream state tying is taken in each data stream of a hand gesture. The basic idea is as the following: first, a whole spatial vector is formed by using all stream vectors. For each hand gesture, its HMM is trained with training samples. After all HMMs have been trained, the observation probability densities in each data stream of all signs are tied with a few probability densities. The advantage of this approach is that the computation time is greatly reduced because the state observation probability density in a whole gesture space is the product of the state observation probability densities of the six streams. And because the state probability in log domain can be computed by summing all stream state probabilities in log domain as the following,

$$\log b_i(x) = \sum_{s=1}^{6} \log b_{is}(x_l)$$

$b_i(x)$ is the *i'th* state observation probability, $b_{is}(x_l)$ is the *s'th data* stream observation probability in the *i'th* state. For each data stream the stream state probabilities are clustered, the stream state probabilities belonging to the same class need to be computed only once. As the number of distinguishable patterns in each data stream is relatively small, for a given observation vector, after these six observation probabilities have been computed, the log likelihood of each sign can be easily gotten by a lookup table and by 5 times addition operations.

In addition, the other advantages of this approach are as follows,

The data stream of left hand and right hand can independently be used to select sign candidates during recognition, for example, the data stream of right-hand position and shape can be used to select the sign candidates without need of computation of all streams.

For the case of continuous sign recognition, even the network topology is linear, because the computation time for the state observation probabilities is relatively small, the probably active sign candidates can be quickly determined.

**4. The algorithm for isolated sign recognition**

Each sign has its own start position, and the start positions of different signs may be different. For isolated sign recognition observation data of several frames near the start position can be used to select the sign candidates, computation load can be greatly reduced in this way.

For the isolated sign language recognition, the algorithm for isolated sign recognition is proposed as the following:

The sign vocabularies are clustered according to parameters in start state of each stream model and its corresponding codebook. Suppose that each stream codebook is $VQ_k$ ($k$=1-6), the sign vocabulary set is *WordSet*. A unique sign subset corresponding to each codeword $i$ in $VQ_k$ is denoted by *SubSet(k,i)*. For different codewords such as i and j, their intersection set is empty, i.e.

$$SubSet(k,i) \cap SubSet(k,j) = \phi$$

and

$$\bigcup_i SubSet(k,i) = WordSet$$

Let the conditional probability of the observation vector $o_k$ with respect to the codeword $i$ and stream $k$ be $p(o_k|k,i)$. Its *posterior* probability can be calculated as the following,

$$p(i|o_k,k) = p(o_k|k,i) / \sum_{j \in VQ_k} p(o_k|k,j)$$

If the *posterior* probability is greater than a threshold, then the sign subset *SubSet(k,i)* is active, otherwise inactive. For each stream observation vector $o_k$ if a codeword is active, then its corresponding sign subset is active. Let all active sign subsets in all stream $k$ be WordCD($k$)($k$=1-6), the active sign subset in all streams will be their intersection set as the following,

$$WordCD = \bigcap_k WordCD(k)$$

Signs in active sign subset *WordCD* will be selected as candidates for further detailed match, and the result is obtained.

Different stream combination coding schemes can be used to further reduce the computation time of the observation probabilities in whole sign space. The procedure can be formed as a hierarchical structure. For example, for left hand data stream when the intersection set of two different sign subsets in two streams is not empty, i.e. $SubSet(k_1,i_1) \cap SubSet(k_2,i_2) \neq \phi$, then the addition operation for signs in the intersection set only needs once to compute the whole observation densities. Based on the same principle, for the combinations over two streams, the similar approach can be obtained.

## 5. Continuous Sign Recognition

### 5.1 Movement epenthesis models

For all possibly reasonable combination of two signs the movement epenthesis models or sign transition models need to be trained, this leads to several sentence level training signs required. In addition, movement epenthesis is not well defined in the sign language books, which also leads to model the movement epenthesis even more difficult. To solve the problem HMMs are used to model the movement epenthesis. The parameters in each HMM of movement epenthesis need to be estimated by sentence level training samples. Because the HMM parameters of signs in a sentence have been estimated by isolated sign training samples, the parameters to be estimated are those of HMMs of movement epenthesis. To estimate these parameters, the HMMs of signs in the sentence and HMMs for movement epenthesis are linked HMM, for example, for a sentence consisting of two signs such as u and v, the whole HMM is shown as Fig.2. The sentence training samples are used to train the parameters in HMM of the movement epenthesis. During training the parameters of signs in the sentence are fixed, only parameters in the sign transition or movement epenthesis HMMs need to be estimated. Note that the transition probability from the last state of *u* to the first state of transition model CD (v|u) is usually set to 1.

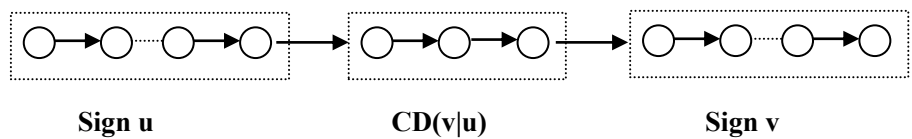

**Sign u**　　　　**CD(v|u)**　　　　**Sign v**

**Figure 2. Context Dependent Modeling**

### 5.2 Tying of sign transition models

Sign transition models have redundancy, and some are impossible combinations between two signs. This kind of sign transition models can be omitted. The possible combinations between *u* and *v* are also too large to be managed. Therefore, model tying for sign transition models is necessary. The similar approach to tying sign models can be used to tie sign transition models.

To further simplify sign transition models, the following approach is proposed. The assumptions are

1. The movement epenthesis between two signs is linearly changed over time.
2. The period of movement epenthesis is relative shorter than that of a sign.

From the first assumption sign transition model can obtained by using linear interpolation, $\lambda_{v|u} = \frac{1}{2}(\lambda_u + \lambda_v)$, where $\lambda_u, \lambda_v$ are the last state model of $u$ and the first state model of $v$. This means that only one state HMM, i.e., GMM, is used to model the sign transition.

From the second assumption, we know that the influence region of the sign transition is much smaller than that of a sign. This assumption is not always true, but for most of sign transitions it is hold.

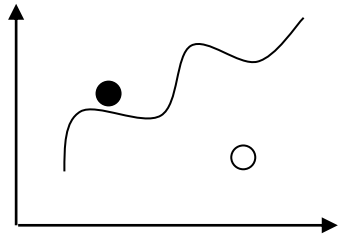

**Figure.3 Linear interpolation for sign transition**

The advantage of this approach is its little computation load. For the case of sentence level training samples available the sign transition models can be further trained with the training samples. For the case of no sentence level training samples available, the average model $\lambda_{v|u} = \frac{1}{2}(\lambda_u + \lambda_v)$ can be used.

### 5.3 Fast matching approach

Each sign has its own trajectory in sign space, if an observation vector was close to the trajectory, the sign would be probably active at that time, otherwise the sign inactive.

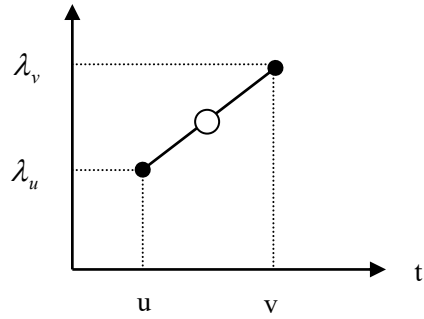

**Figure.4 A trajectory of a sign in sign space**

For the observation vector of each frame how to judge whether a sign is active becomes very important to speeding up the recognition process. If only a small fraction of signs is active at a frame, the most probably active signs are what are active at the previous frame due to the continuity of gestures. Only these active signs need to be further searched at the next frame, thus a large mount of computation load is reduced.

The approach to selecting the active signs at a frame is as the following: let $U_i$ be a basic sign unit not including sign transition unit. If a state belongs to the state set of units $U_i$, then denote $j \in U_i$. For each frame the active score of stream $s$ of the unit is computed as the following,

$$P_{U_i}(t,s) = \max_{j \in U_i} \log b_j(s, O_t)$$

$b_j(s, O_t)$ is the state observation probability, and $P_{U_i}(t,s)$ denotes the active score. This score can be used as an active measure to order the recognition units. A threshold also can be used to select the active units. The active unit will be further searched at the next frame.

The following problems may exist in this approach:
Similar with other fast match approaches, this approach may cause the pruning errors, because if one unit is incorrectly pruned at some frame, the errors may not be recovered. The algorithm needs computing all state probabilities of all basic sign units. The computation load is relatively large for a fast match approach.

The approach to selecting active sign transition units is as the following: to reduce the computation time in the sign transition units, Rich Get Richer strategy is used [26]. For sign transition unit v|u, let *pu* be the last state probability of the unit *u*, let *pv* be the first state probability of the unit *v*, the

look-ahead score of sign transition unit v|u is approximated by *pv|u=(logpu+logv)/2+lookahead* (v), where the *lookahead*(v) score is the average matching score of the observation vectors at the following frames t+1,t+2,t+3 to the first state of unit *v*. The look ahead score can be used as an active measure and to order sign transition units. The active sign transition unit is further extended.

## 6. The search algorithm

A recursive transition network composed of the states of HMMs is used to model sentence level sign. Viterbi algorithm is employed for decoding. In Viterbi beam searches only the hypothesis whose likelihood falls within a beam of the most likely hypothesis is considered for further growth. The best beam size is determined empirically. By expanding the network to include an explicit arc from the end of each sign to the start of the next, the bigram language model has been incorporated to improve recognition accuracy in our system. To speed up the decoding process, the fast match approach discussed in previous section and the following pruning approach are used.

To conserve the computing and memory resources, it is imperative to prune the low-scoring partial paths. Two different beams are used, one is at state level, and the other is at sign level. The beam width at each level is determined empirically, and the beam threshold is computed with respect to the best path scoring at that level. For sign-to-sign transitions, if the cumulative score for the partial hypothesis in the last state of a sign exceeds the beam, no transitions are computed from this sign.

## 7. Experiment

The baud rate for both CyberGlove and 3-D tracker is set 38400. The number of states in HMM of each sign is 3 0r 5, which is determined by an adaptive approach. The raw gesture data, which in our case are values of 18-joint angles collected from the Cyberglove for each hand, the range of each angle value is within 0-255. For two hands, they are formed as a 48-dimensional vector appended with hand and position and orientation features. The dynamic range of each component is different. Each component value is normalized to ensure its dynamic range is 0-1 [24].

The HMM structure for each sign is left to right without skip. The hardware environment is Pentium III 450Hz.

### 7.1  Isolated sign recognition evaluation

For the case of isolated sign recognition 5177 signs in Chinese sign language were used as evaluation vocabularies. Each sign was performed 5 times by a sign language teacher. 4 times were used for training and one for test. Using the approach of cross validation test, the test times for each sign is 5. For different numbers of different patterns in each stream, the off-line recognition rates are listed in the Table 1. Where Lp, Lo, Ls, Rp, Ro, Rs, are the number of different patterns in data stream of left/right hand position, left/right hand orientation, left/right hand shape, respectively. The number of states in HMM is 3. When The number of states in HMM is 5, the results are listed in the Table 2. This shows that 3 states are better than the 5 states considering the computation load and recognition accuracy. Therefore 3 states for each HMM are used in the continuous recognition.

Table 1. The Recognition rates ( 3 states)

| Lp | Lo | Ls | Rp | Ro | Rs | Accuracy |
|---|---|---|---|---|---|---|
| 64 | 64 | 64 | 64 | 64 | 64 | 91.6% |
| 128 | 128 | 128 | 128 | 128 | 128 | 93.4% |
| 256 | 256 | 256 | 256 | 256 | 256 | 94.8% |
| 512 | 512 | 512 | 512 | 512 | 512 | 95.5% |
| 1024 | 1024 | 1024 | 1024 | 1024 | 1024 | 95.8% |

Table 2. The Recognition rates (5 states)

| Lp | Lo | Ls | Rp | Ro | Rs | Accuracy |
|---|---|---|---|---|---|---|
| 64 | 64 | 64 | 64 | 64 | 64 | 91.9% |
| 128 | 128 | 128 | 128 | 128 | 128 | 93.4% |
| 256 | 256 | 256 | 256 | 256 | 256 | 94.6% |
| 512 | 512 | 512 | 512 | 512 | 512 | 95.0% |
| 1024 | 1024 | 1024 | 1024 | 1024 | 1024 | 95.4% |

For online test because the property of Viterbi decoding is time synchronous and the recognition time is less than that needed for performing a sign, once a sign is stopped, the system will report the recognition results without any noticeable delay.

### 7.2 Continuous sign recognition evaluation

For the case of continuous recognition, the database of gestures consists of 5177 signs and 500 sentences. In general, each sentence consists of 2 to 15 signs. No intentional pauses were placed between signs within a sentence. Sign transition model discussed in section 5.2 was used to model the movement epenthesis.

To test the recognition performance at sentence level, one test was carried out described as the following. When 500 sentences were not used for any portion of the training. The 5177 words are used as basic units. Within 500 sentences, 264 sentences can be correctly recognized, the left 236 sentences have deletion (D), insertion (I), and substitution (S) errors, D=186, I=302, S=332, N=5162, N denotes the total number of signs in the test set, the word correct rate is 84.1%.

This shows that the movement epenthesis has affected on recognition performance at sentence level. To consider this effect, the sign transition HMMs were trained by the sentence level training samples. For the left 264 sentences, the training procedure for sign transition models was used for each of sentence. In the collected sentence samples, four in five are used for training and one for test. The word correct rate is 91.26%, where D=98, I=182, S=171, N=5162. The accuracy measure is calculated by subtracting the number of deletions, insertion, substitution, and errors from the total number of signs and divided by the total number of signs. The result shows that sign transition models are necessary for sentence level recognition. The recognition speed is about 2 times of real-time. When the number of sign transition models is set to 3 for all sign transition model, the word correct rate is 91.4%, where D=97, I=180, S=167, N=5162. The recognition rates are summarized in the Table 3.

Table 3. The word correct rates for different number of states in sign transition models

| The number of states in sign transition model | Recognition Errors | Word Correct Rate |
|---|---|---|
| 1 | D=98   I=182   S=171 | 91.26% |
| 3 | D=97   I=180   S=167 | 91.40% |

## 8. Conclusion

We have presented a framework for large vocabulary sign recognition by stream state tying. For isolated sign recognition, we propose an algorithm for large vocabulary isolated sign recognition, the algorithm is 5-85 times faster than the approach without tying. As for the continuous sign recognition we propose a simplified sign transition model and a fast-matching approach to speeding up the decoding speed. Experiments have shown that these techniques are powerful for sign language recognition considering the problem of scalability.

**Acknowledgements**

This work was supported in part by Chinese NSF, Chinese national 863 Hi-tech development program intelligent computing system subject and Kuancheng Wang's postdoctoral award of Chinese Academy of Sciences. The research was conducted while the first author held a postdoctoral position from 1999 to 2001.